\documentclass[english,floatfix,twocolumn, aps, prl, reprint, superscriptaddress, longbibliography]{revtex4-2}

\usepackage{amsmath, amssymb}
\usepackage{graphicx}
\usepackage{booktabs, footnote}
\usepackage{physics}
\usepackage{xcolor}

\usepackage{hyperref}
\hypersetup{colorlinks=true, citecolor=blue, urlcolor=blue, linkcolor=blue, breaklinks=false}

\begin{document}

\title{Observation of an anomaly in the statistics of Kibble-Zurek defects}

\author{Jan Balewski$^\dagger$}
\thanks{These authors contributed equally.\\
$^\dagger$Email address:
\href{mailto:balewski@lbl.gov}{balewski@lbl.gov};
\href{mailto:alehud@bu.edu}{alehud@bu.edu};
\href{mailto:kklymko@lbl.gov}{kklymko@lbl.gov};
\href{mailto:dcamps@lbl.gov}{dcamps@lbl.gov};
\href{mailto:mkornjaca@quera.com}{mkornjaca@quera.com}.
}
~\affiliation{National Energy Research Scientific Computing Center, Lawrence Berkeley National Laboratory, Berkeley, CA 94720, USA}
\author{Alexey Khudorozhkov$^\dagger$}
\thanks{These authors contributed equally.\\
$^\dagger$Email address:
\href{mailto:balewski@lbl.gov}{balewski@lbl.gov};
\href{mailto:alehud@bu.edu}{alehud@bu.edu};
\href{mailto:kklymko@lbl.gov}{kklymko@lbl.gov};
\href{mailto:dcamps@lbl.gov}{dcamps@lbl.gov};
\href{mailto:mkornjaca@quera.com}{mkornjaca@quera.com}.
}
~\affiliation{QuEra Computing Inc., 1284 Soldiers Field Road, Boston, MA, 02135, USA}
~\affiliation{Department of Physics, Boston University, Boston, MA 02215, USA}

\author{Siva Darbha}
~\affiliation{National Energy Research Scientific Computing Center, Lawrence Berkeley National Laboratory, Berkeley, CA 94720, USA}

\author{Omar A. Ashour}
~\affiliation{National Energy Research Scientific Computing Center, Lawrence Berkeley National Laboratory, Berkeley, CA 94720, USA}
~\affiliation{Department of Physics, University of California, Berkeley, California 94720, USA}

\author{Fangli Liu}
~\affiliation{QuEra Computing Inc., 1284 Soldiers Field Road, Boston, MA, 02135, USA}

\author{Ermal Rrapaj}
~\affiliation{National Energy Research Scientific Computing Center, Lawrence Berkeley National Laboratory, Berkeley, CA 94720, USA}

\author{Sheng-Tao Wang}
~\affiliation{QuEra Computing Inc., 1284 Soldiers Field Road, Boston, MA, 02135, USA}

\author{Pedro L. S. Lopes}
~\affiliation{QuEra Computing Inc., 1284 Soldiers Field Road, Boston, MA, 02135, USA}

\author{Katherine Klymko$^\dagger$}
~\affiliation{National Energy Research Scientific Computing Center, Lawrence Berkeley National Laboratory, Berkeley, CA 94720, USA}

\author{Milan Kornja\v ca$^\dagger$}
~\affiliation{QuEra Computing Inc., 1284 Soldiers Field Road, Boston, MA, 02135, USA}

\author{Daan Camps$^\dagger$}
~\affiliation{National Energy Research Scientific Computing Center, Lawrence Berkeley National Laboratory, Berkeley, CA 94720, USA}

\date{\today}

\begin{abstract}
The Kibble-Zurek mechanism quantifies defect formation during finite-rate passage across a continuous phase transition, providing key insights into universality in quantum many-body systems. We explore counting statistics of defects in finite-rate ramp experiments on long 1D Rydberg atom chains. The experiments reveal an anomaly in the defect number distribution at long ramp times, challenging the hypothesis of defect formation through independent domain mergers. Numerical simulations reproduce the anomaly and support an interpretation in terms of post-critical coarsening during the ramp, while prepare-and-hold experiments reveal no subsequent secular growth of the measured correlations. Our results show the limits of classical or integrable system descriptions of Kibble-Zurek defect formation for interacting quantum systems.
\end{abstract}

\maketitle

The Kibble-Zurek (KZ)~\cite{Kibble1976, Zurek1985, Dziarmaga2010} mechanism describes universal topological defect dynamics for a system driven across a continuous phase transition at a finite rate. As the control parameter approaches criticality, the diverging relaxation time causes the dynamics to ``freeze out'', producing a characteristic scaling $\langle D \rangle \propto \Gamma^{\mu}$ of the mean defect number $\langle D \rangle $ with the ramp rate $\Gamma$, where $\mu$ is determined solely by equilibrium critical exponents. The KZ mechanism has been verified experimentally across systems ranging from liquid crystals to superconductors to ultracold gases, establishing a central paradigm in nonequilibrium statistical mechanics~\cite{Chuang1991, Monaco2009, Weiler2008, Pyka2013, Ulm2013, Xu2016, Rysti2021, Bando2020, Du2023}. In quantum systems, Rydberg atom arrays have emerged as powerful probes of phase-transition dynamics: experiments have observed universal KZ scaling in one- and two-dimensional Ising transitions, explored real-time dynamics across quantum critical points and phases, and revealed non-critical coarsening that persists long after crossing a transition~\cite{Bernien2017, Keesling2019, Ebadi2021, Manovitz2024, Hirsbrunner2025, Darbha2025}. These experiments have extended KZ studies from measuring the average defect number scaling towards resolving defect correlations in a tunable setting.

\begin{figure*}[!htb]
\centering
\includegraphics[width=0.9\textwidth]{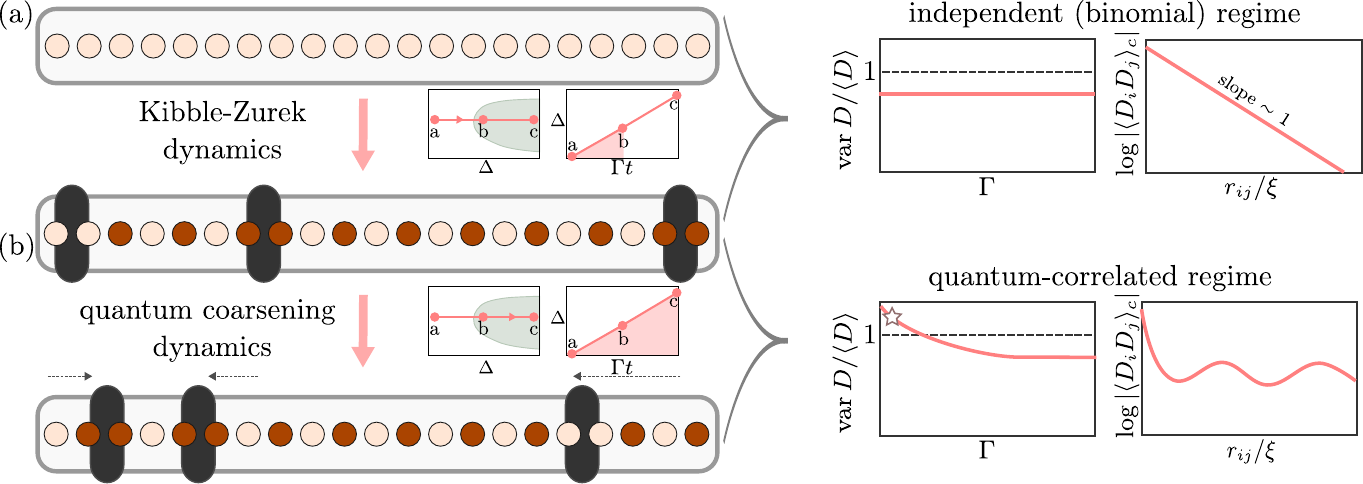}
\caption{\textbf{Kibble-Zurek defect formation and statistics anomaly.} (a) Schematic of a Rydberg atom array with Ising spins (orange/brown dots). As the system is driven from a paramagnetic phase `a' through the vicinity of a critical point `b' into the antiferromagnetic phase, pre-domains of antiferromagnetic order develop on a correlation length ($\xi$) scale. The mergers between the domains result in the formation (black) of the domain wall defect at random. This results in the Poisson-binomial defect counting statistics at time point `b', with variance proportional to and smaller than the mean, and uncorrelated defects (top right). (b) For slow driving rates $\Gamma$, we propose that defects become correlated through coarsening dynamics deep within the antiferromagnetic phase lobe, `c', resulting in a defect counting statistics anomaly (bottom right).
}
\vspace{-12pt}
\label{fig:predomains}
\end{figure*}

While Kibble-Zurek scaling characterizes the mean defect density,
the statistics of defect formation can encode richer information. Recent theoretical studies have proposed, however, that all cumulants of the defect counting distribution scale universally with the same KZ exponent~\cite{delCampo2018, GomezRuiz2020, delCampo2021, GomezRuiz2022, delCampo2022, delCampo2023}. The results follow from an ``independent-domain-merger'' picture [Fig.~\ref{fig:predomains}(a)]: as the system approaches criticality, pre-domains nucleate on length scales set by the Kibble-Zurek correlation length $\xi$. These domains merge without a topological defect only when their ordering is compatible. Assuming that the merges are independent and random, the defect formation is naturally modeled as a Poisson-binomial process. As a direct consequence, all cumulants inherit the same Kibble-Zurek scaling, while the variance satisfies the Poisson-binomial bound $\operatorname{Var}(D)\leq\langle D\rangle$.

This hypothesis was inspired by studies in integrable models~\cite{delCampo2018}, where free quasiparticles enforce the independent-domain-merger picture, and has proven remarkably successful. Experiments on quantum annealers and superconducting qubit processors have reported defect statistics consistent with the theoretical expectations for integrable models, reinforcing the view that defect formation reflects uncorrelated domain mergers~\cite{Bando2020, Cui2020, Kiss2025}. Yet these successes raise a fundamental question: does the counting statistics universality based on the independent-domains assumption remain valid in more ubiquitous non-integrable, short-ranged, strongly-interacting quantum systems~\cite{Kolodrubetz2012a,Gherardini2024, Roychowdhury2021, Mattioli2025}?

In this letter, we find that the universality of defect statistics is violated in a non-integrable quantum system. Using Rydberg chains with up to 58 atoms, we measure the full counting statistics of KZ defects across the $\mathbb{Z}_2$ phase transition, observing timescale-dependent behavior. At higher ramp rates, the first two measured moments are compatible, within experimental uncertainty, with the independent-defect variance bound. At lower rates, we observe a clear violation of this bound: the variance exceeds the mean, and the defect distribution becomes super-Poissonian. Exact-diagonalization simulations reproduce this anomaly and reveal a connection with long-range defect-defect correlations. We interpret the anomaly in defect statistics, and the development of defect correlations as induced by post-critical coarsening~\cite{Bray1994, Bray2003, Manovitz2024, Samajdar2024} [Fig.~\ref{fig:predomains}(b)]. Finally, prepare-and-hold experiments reveal no subsequent secular
growth of the measured correlations, finding a phenomenology in sharp contrast to recent two-dimensional results~\cite{Manovitz2024}, thus opening a new line of inquiry.

We explore defect statistics using the Aquila~\cite{Wurtz2023, bloqadepython:2024} Rydberg atom quantum simulator, which operates the coherent dynamics of Rubidium arrays under the Rydberg Hamiltonian
\begin{equation}
\begin{split}
\frac{\mathcal{H}(t)}{\hbar} &= \frac{\Omega(t)}{2} \sum_j \left(  \outerproduct{g_j}{r_j} +  \outerproduct{r_j}{g_j} \right) \\
&~~~~ - \Delta(t) \sum_j n_j + C_6 \sum_{j < k}\frac{n_j n_k}{|\Vec{r}_j - \Vec{r}_k|^6} \, ,
\label{eq:ham}
\end{split}
\end{equation}
with $j$ indexing the atoms arranged in a chain with nearest neighbor distance $a$. $\Omega(t)$ and $\Delta(t)$ are, respectively, the Rabi frequency and detuning coupling the ground states $\ket{g_j}$ and highly-excited Rydberg states $\ket{r_j}$, with excited state population $n_j = \outerproduct{r_j}{r_j}$. The strong Van der Waals interactions between excited atoms lead to the effective Rydberg blockade, whereby simultaneous excitations of atoms within the radius $R_b = (C_6 / \Omega)^{1/6}$ are strongly suppressed. 

The Rydberg Hamiltonian is non-integrable, with its ground state phase diagram studied extensively in experiments~\cite{Bernien2017, Keesling2019, Zhang2025}. We focus on the first-neighbor blockade regime $R_b / a \in [1, 2]$. At low detuning, the ground state of the system is paramagnetic, adiabatically connected to the all-ground product state. At high detuning, the system is in an antiferromagnetic phase, adiabatically connected to the product state with the maximum number of Rydberg excitations consistent with the first neighbor blockade. A quantum critical transition in the $\mathbb{Z}_2$ universality class separates the two phases~\cite{Sachdev2011, Sengupta2004}, characterized by $\mu=0.5$ Kibble-Zurek scaling exponent, probed previously in Ref.~\cite{Keesling2019}. This transition is used in our experiments to probe the counting statistics of defects and test the independent-defect theory in the non-integrable setting.

\begin{figure}[tb]
\centering
\includegraphics[width=1.0\columnwidth]{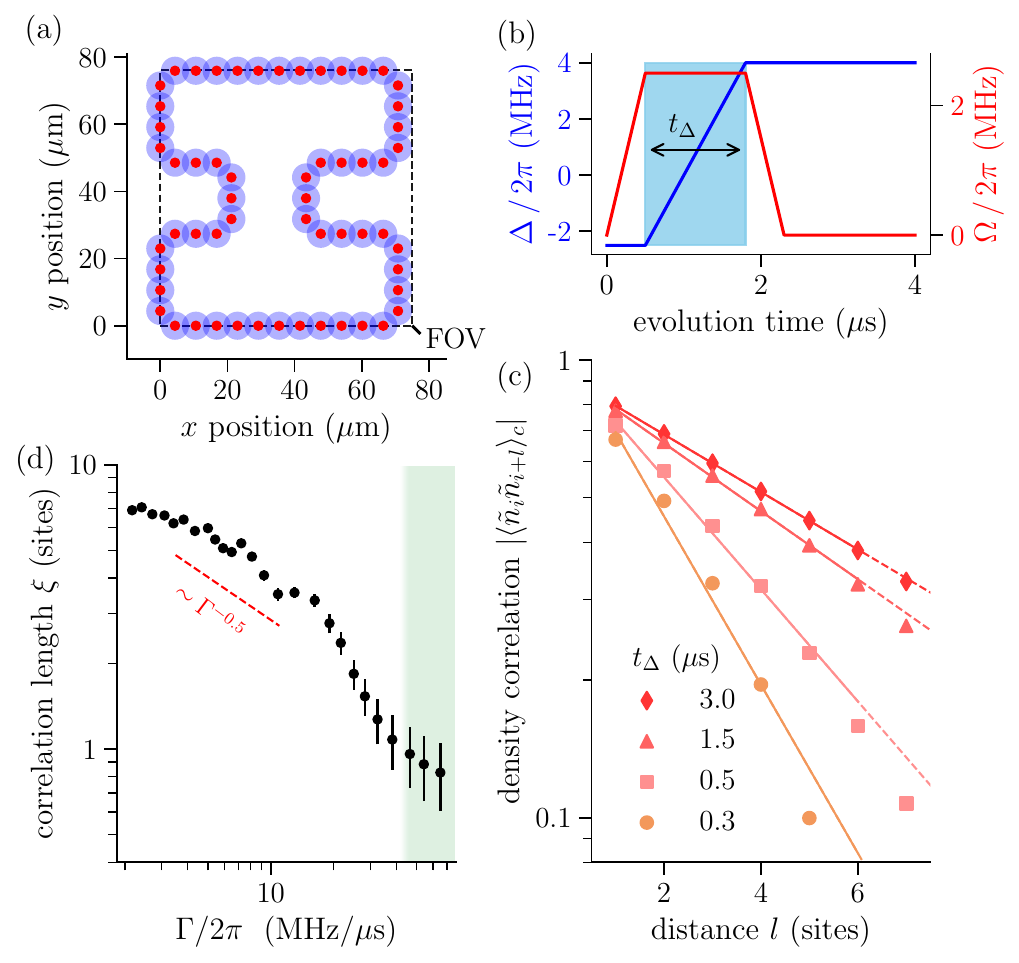}
\caption{\textbf{Experimental setup and correlation length measurement.} 
(a) The 58 atom geometry chosen to fit in the Aquila simulator~\cite{Wurtz2023} field of view. Red circles indicate the atom positions separated by $a=6.2~ \mu \mathrm{m}$, and semi-transparent blue circles show blockade radii $R_b / a = 1.35$. 
(b) The Rabi frequency $\Omega(t)$ and detuning $\Delta(t)$ drives used to cross the transition into the antiferromagnetic regime. The blue shaded region shows the variable detuning ramp time $t_\Delta \in [0,3]\mu$s. 
(c) The two-point Rydberg density correlations $|\langle \tilde{n}_i \tilde{n}_{i+l} \rangle_c|$ ($\tilde{n}=n-1/2$) as a function of distance $l$ for different ramp times. The solid lines show exponential fits over $l \in [1,6]$ used to extract correlation lengths. 
(d) The defect correlation length vs the quench rate $\Gamma$. For intermediate quench rates, the behavior approaches the expected $\mu=0.5$ Kibble-Zurek scaling (red line). 
Fast quench rates, defined by $\langle D \rangle \gtrsim L/4$, are shown by the green region. Slow quench rates, defined by $\langle D \rangle < 1$, fall below the $\Gamma$-axis limits. 
}\vspace{-12pt}
\label{fig:KZscaling}
\end{figure}

To facilitate experiments at scale, we arrange 58 atoms in a loop-chain configuration [Fig.~\ref{fig:KZscaling}(a)] consistent with experimental spacing constraints~\cite{Wurtz2023}. We perform a series of finite-rate ramp preparation experiments characterized by the Rabi drive $\Omega(t)$ with its maximal value yielding $R_b / a = 1.35$, while linearly ramping the detuning $\Delta(t)$ from large negative values, through the $\mathbb{Z}_2$ transition~\cite{Bernien2017} [Fig.~\ref{fig:KZscaling}(b)], to a constant final value well within the antiferromagnetic phase. The detuning ramp duration, $t_\Delta$, controls the rate, $\Gamma = (\Delta_{\mathrm{max}} - \Delta_{\mathrm{min}}) / t_{\Delta}$, at which the transition is crossed. The experimental constraints enable a sizable dynamic range, $\Gamma/2\pi \in (2, 70)$~MHz/$\mu$s. For each value of $t_\Delta$, we collect $2000$ measurement shots, retaining approximately $75\%$ after postselecting for defectless initial atoms placement.

We first measure the correlation length, $\xi$, of the ramp-prepared states and its dependence on the ramping rate by estimating the expectation value of the density-density correlations and fitting them to exponential decay, $|\langle n_i n_{i+l} \rangle| \propto e^{-l/\xi}$. The correlation length fits for several characteristic ramping rate values are presented in Fig.~\ref{fig:KZscaling}(c), showing a clear trend towards higher correlation length with slower ramps. Even though the readout error, significant in this analog simulator~\cite{Wurtz2023}, affects the correlation functions, the effect is uniform across displacement $l$, making correlation length estimation based on fitting robust~\cite{SM}.

\begin{figure}[tb]
\centering
\includegraphics[width=1.0\columnwidth]{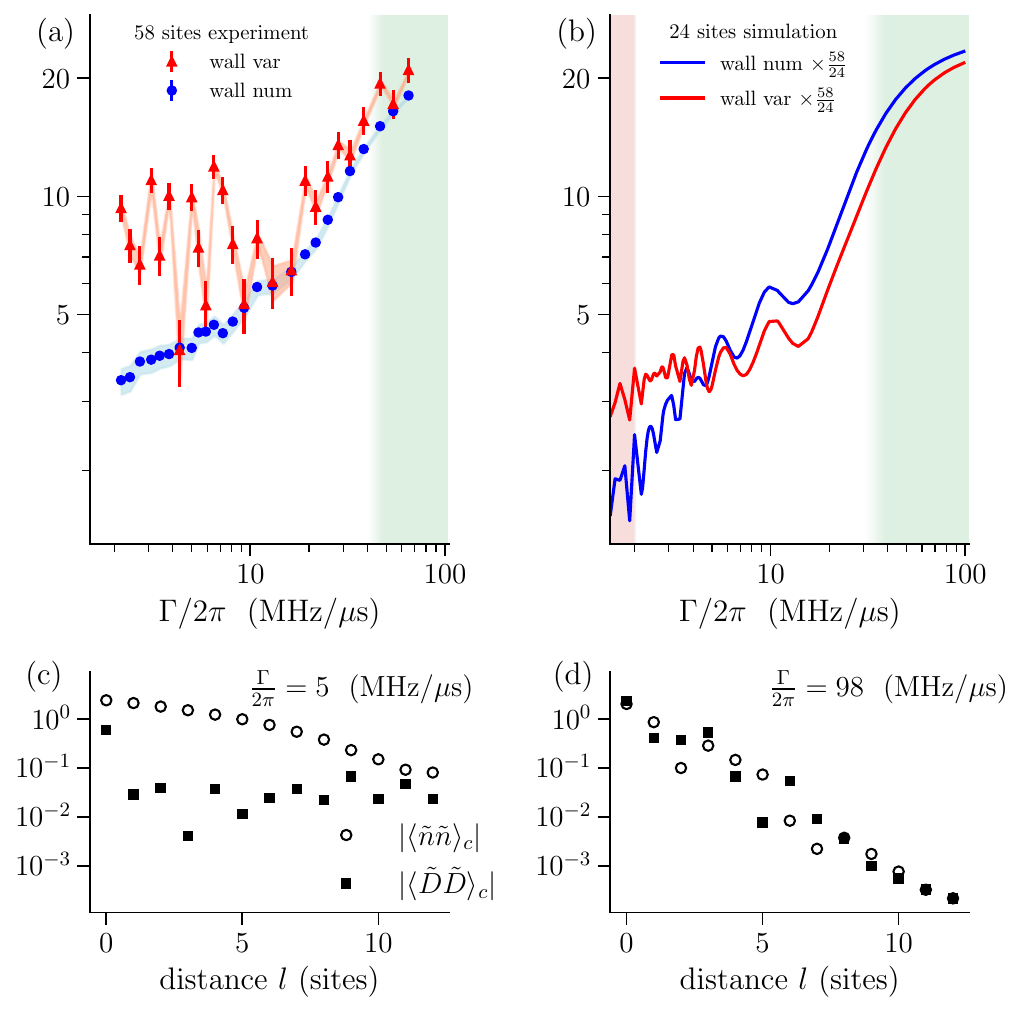}
\caption{\textbf{Anomaly in the defect counting statistics.} 
Domain wall mean (blue) and variance (red) vs the quench rate $\Gamma$ from (a) experiments and (b) numerical simulations rescaled by the size ratio $58/24$.
Vertical error bars show one-standard-error uncertainties of the zero-noise-extrapolated estimates, obtained by propagating the statistical uncertainties through the extrapolation procedure. Shaded bands show the maximum systematic changes in the estimates when the calibrated readout-error probabilities are varied within their quoted uncertainties~\cite{SM}. The fast ramp regime, defined by $\langle D \rangle \gtrsim L/4$, is shown in green, while the slow ramp regime, defined by $\langle D \rangle < 1$, is shown in red. 
In both experiments and numerics, we observe a salient regime at lower quench rates where the defect variance exceeds the mean. The discrepancy is associated with the development of non-trivial defect correlations. In numerics, we compare the connected correlators of the density-density and defect-defect observables: (c) for slow ramps, short-range domain wall correlations are suppressed, coinciding with a statistics anomaly; (d) at faster ramps, domain wall correlations simply follow the correlation length determined by density correlations.
} 
\vspace{-12pt}
\label{fig:KZanomaly}
\end{figure}

To distinguish dynamical ramp regimes without reference to the counting-statistics anomaly discussed below, we use the mean domain-wall separation, $\ell_D = L / \langle D \rangle$. We define the fast-quench regime by $\langle D \rangle \gtrsim L/4$, or equivalently $\ell_D \lesssim 4$ lattice sites (two antiferromagnetic unit cells). In this regime the characteristic domain size is microscopic, the measured correlation length approaches its short-distance saturation value, and the critical region is traversed too rapidly for long-range correlations to develop. At the opposite end, we define the finite-size regime by $\langle D \rangle < 1$, where the expected number of defects is below unity and the thermodynamic freeze-out picture is cut off by the finite-size gap. The intermediate-rate regime is defined by $1 \lesssim \langle D \rangle \lesssim L/4$. The corresponding boundaries, $\Gamma_{\mathrm f}$ and $\Gamma_{\mathrm s}$, are indicated in Figs.~\ref{fig:KZscaling}(d) and \ref{fig:KZanomaly}(a,b) where applicable.

Within the intermediate-rate window, the measured correlation length is consistent with the expected Kibble-Zurek power law, $\xi \sim \Gamma^{-0.5}$, reproducing the behavior reported in Ref.~\cite{Keesling2019}. At higher rates, $\xi$ saturates at a microscopic value, marking the fast-quench regime. Importantly, all experimental data satisfy $\langle D \rangle > 1$; the experiment therefore does not enter the finite-size regime~\cite{Dziarmaga2010,Kolodrubetz2012b,Zhang2025b}. That regime is accessed only by the lowest-rate numerical simulations discussed in the Supplementary Materials~\cite{SM}.

The counting statistics of defect formation is logically distinct from the Kibble-Zurek power law. If defect formation is described by independent Bernoulli events, as proposed in Refs.~\cite{delCampo2018, GomezRuiz2020}, each with probability $p_\alpha$, then $\langle D \rangle = \sum_\alpha p_\alpha$ and $\operatorname{Var}(D)= \sum_\alpha p_\alpha (1-p_\alpha) \leq \langle D \rangle$ irrespective of the detailed dependence of $p_\alpha$ on the ramp rate. Kibble-Zurek scaling further constrains that rate dependence within the critical scaling window, leading to the same KZ exponent for all higher order cumulants of defect counting statistics. However, a breakdown of the Kibble-Zurek power law does not by itself imply a breakdown of statistical independence. This distinction is not explicit in integrable or classical models, i.e. effectively independent-defect ones, for which Poisson-binomial statistics have been obtained over broad ranges of excitation densities~\cite{delCampo2018,GomezRuiz2020,Cui2020,delCampo2023}. These results do not, however, provide an a priori argument that the independent defect picture remains valid in a non-integrable interacting system. 

Here, we test the independent-defect hypothesis through its most direct moment constraint, $\operatorname{Var}(D) \leq \langle D \rangle$. Experimentally, the number of defects is obtained from nearest-neighbor correlations that identify the two types of antiferromagnetic domain walls:
\begin{equation}
\langle D \rangle = \sum_i \langle D_i \rangle = \sum _i \langle n_i n_{i+1} + (1-n_i) (1-n_{i+1}) \rangle \, .
\label{eq:D_operator}
\end{equation}
The variance follows from four-point density correlators $\operatorname{var}{D} = \langle D^2 \rangle - \langle D \rangle^2$. In contrast to the readout-robust correlation length, the defect number and variance are significantly affected by the readout error, which is skewed $\sim 5\%$ towards $\ket{r_j} \rightarrow \ket{g_j}$ misdetection. To recover reliable estimates of defect distribution moments, we carefully calibrate the readout error~\cite{Wurtz2023} during our data collection and apply zero-noise extrapolation during postprocessing~\cite{SM}. Statistical uncertainties are propagated through the readout
mitigation, while sensitivity to the calibrated readout-error probabilities is shown separately as systematic uncertainty.

Our main experimental results are summarized in Fig.~\ref{fig:KZanomaly}(a). The mean domain-wall density increases with ramp rate, consistent with $\langle D\rangle/L\propto\xi^{-1}$. At higher ramp rates, the measured variance and mean are compatible, within experimental uncertainty, with the independent-defect bound $\operatorname{Var}(D)\leq\langle D\rangle$. This compatibility does not establish the full Poisson-binomial distribution or resolve a distinctly sub-Poissonian experimental regime. At lower rates, the variance clearly exceeds the mean, reaching $\operatorname{Var}(D)/\langle D\rangle\simeq3$ at the lowest experimentally accessible rates. We use $\Gamma_a/2\pi\sim10~\mathrm{MHz}/\mu\mathrm{s}$ to denote the approximate experimental scale over which this change becomes apparent, rather than a sharply resolved crossing. The violation occurs within the intermediate-rate regime, while $\langle D\rangle>1$, and is therefore not confined to the finite-size slow-ramp limit.

To determine whether the many-body dynamics can produce the anomaly in the absence of experimental noise, we perform exact-diagonalization simulations~\cite{BloqadeJulia:2023} of our ramp protocol with up to 24 sites~\cite{SM}. The results in Fig.~\ref{fig:KZanomaly}(b) reproduce the qualitative change from a variance smaller than the mean at higher rates to a variance exceeding the mean at lower rates. The numerical crossing occurs near $\Gamma/2\pi\approx5~\mathrm{MHz}/\mu\mathrm{s}$, below the approximate experimental scale of $10~\mathrm{MHz}/\mu\mathrm{s}$. Since the experimental crossing is not sharply resolved, we do not claim quantitative agreement between these rates. Differences in system size, the perfect-blockade approximation~\cite{SM}, and experimental imperfections may contribute to the offset, but their individual contributions are not quantitatively isolated here. Crucially, this anomaly occurs while $\langle D\rangle>1$. Only at still lower rates~\cite{SM}, $\Gamma < \Gamma_{\mathrm s}$, do the simulations enter the finite-size regime defined by $\langle D \rangle < 1$. Finite-size effects may therefore influence the ultimate low-rate saturation, but they are not required for the onset of anomalous statistics in numerics.

The simulation permits a finer look into defect-defect correlations [see Fig.~\ref{fig:KZanomaly}(c,d)], which is hindered in experiment by prohibitive shot requirements. Tellingly, the defect-defect correlation behavior is distinct in the two regimes. At higher ramp rates, the magnitude and decay length of the defect correlator approximately track those of the density correlator, consistent with defects inherited from correlation-length-sized domains. In the anomalous-statistics regime, the defect correlator instead develops a spatial profile that is not inherited from the density correlations. In particular, the short-distance pair probability is reduced relative to the longer-distance correlation profile: conditioning on the presence of one wall produces a weaker enhancement of the probability of another wall nearby than at larger separations.  Although the defect correlator is of order $10^{-2}$, its natural scale at low defect density is $(\langle D \rangle/L)^2$,; the corresponding relative modulation of the pair probability can therefore be substantial. The observed profile provides a quantitative constraint on statistical models of the defects, while its microscopic dynamical origin remains unresolved.

The anomaly in the defect statistics calls for a mechanism that produces correlated defects. We conjecture that this mechanism is related to non-critical coarsening dynamics that apply after the critical point is crossed~\cite{Bray1994, Bray2003, Samajdar2024}. At higher ramp rates, the shorter time spent in the ordered phase may limit post-critical reorganization, leaving the final statistics compatible with the independent-defect bound. At lower rates, the additional time spent in the ordered phase allows non-integrable dynamics to rearrange and correlate the initially generated defects, potentially invalidating the independent-defect description [Fig.~\ref{fig:predomains}(b)].

Such interaction-induced post-critical reorganization is not captured by the factorized independent-mode treatments underlying earlier counting-statistics predictions~\cite{delCampo2018}, where the evolution proceeds in decoupled quasiparticle sectors. The non-integrable Rydberg Hamiltonian permits these interacting
processes, but non-integrability alone does not specify a unique relaxation time. Within the proposed coarsening picture, the crossover would reflect a competition between the time available for evolution in the ordered phase and the timescales for defect rearrangement. Slower detuning ramps increase this available time, while the Rabi coupling, detuning, interactions, and prepared
defect density determine which rearrangement processes are
accessible. The crossover would therefore depend on both the microscopic dynamics and the ramp protocol, rather than being fixed solely by the equilibrium critical exponents.

We note, however, two alternative mechanisms that could also produce the anomaly, but are inconsistent with our observations. First, long-range interactions are known to introduce an anomaly in the defect distributions, with no classical analog~\cite{Gherardini2024}. In particular, long-range interactions, which decay more slowly than $r^{-d}$ for system dimension $d$, directly correlate defects during their formation. However, this does not apply to our $d=1$ experiments. Second, the anomaly could arise classically in an independent-defect framework for an even-Poisson distribution, stemming from the constraint on pairwise defect formation~\cite{delCampo2021}. Such a constraint commonly arises from topology and is indeed present in our system with periodic boundary conditions~\cite{SM}. Our experimental and numerical results, however, disagree with the implications of even-Poisson theory. This theory predicts that the sizable anomaly $\operatorname{var}{D} > \langle D \rangle$ for the slow ramps regime is followed by an increasing (doubling) slope of both the variance and the mean, while the opposite (flattening and slowdown of decay) is observed in the experiment and simulations. Additionally, it predicts that $\operatorname{var}{D} > \langle D \rangle$ is valid at all ramp speeds, with exponential convergence between the variance and mean for fast ramps. This prediction is also contradicted by the numerical variance being smaller than the mean at higher rates, whereas the experimental uncertainties do not resolve this distinction.

\begin{figure}[tb]
\centering
\includegraphics[width=1.0\columnwidth]{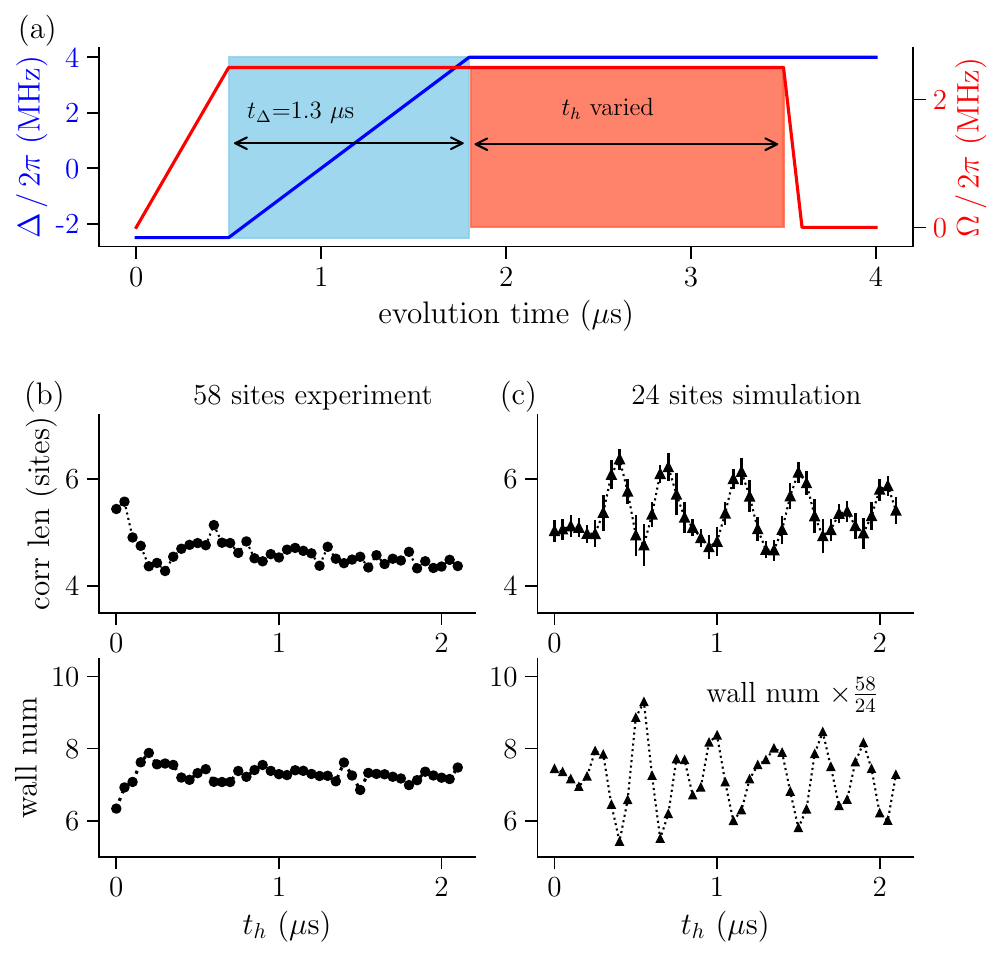}
\caption{\textbf{Hold protocol and dynamics.} (a) By appending the time constant protocol for a hold time $t_h$ after the ramp, we probe the non-critical dynamics at fixed energy. Both in (b) experiment and (c) simulations, correlation lengths and domain wall numbers are stable in time, indicating a freeze out of the coarsening dynamics. Simulations further recover oscillations absent in experiments due to dephasing.}
\vspace{-12pt}
\label{fig:coarsening}
\end{figure}

Finally, we devise and perform experiments to further probe and control the non-critical coarsening dynamics. The hold protocol, shown in Fig.~\ref{fig:coarsening}(a), consists of a fixed, intermediate speed ramp followed by a constant Hamiltonian evolution for different hold times ($t_h$). This protocol is equivalent to the one recently used to probe coarsening dynamics of the two-dimensional Rydberg Hamiltonian~\cite{Manovitz2024}. The correlation length and defect number are measured as a function of hold time, with experimental and simulated results presented in Figs.~\ref {fig:coarsening}(b) and (c).

The hold dynamics exhibit constant correlation length and defect number on average in both the experiments and simulations. The absence of secular growth during the final hold does not exclude rearrangement during the ramp, when the instantaneous Hamiltonian and the accessible processes differ. Furthermore, the numerics resolve the secondary effect of coherent oscillations in the correlation length, with the period governed by the ground state energy gap~\cite{SM}, akin to the observations in two-dimensions~\cite{Manovitz2024}. These oscillations are absent in the experiment, likely due to dephasing~\cite{Wurtz2023, Darbha2025}. The freezing of correlation length through coarsening is a surprising result, in sharp contrast with the observations in two-dimensions of diffusive growth, $\xi \propto t^{0.5}$~\cite{Manovitz2024}. It showcases further control over coarsening dynamics that could be employed in future experiments to tune the appearance and properties of the defect statistics anomaly.

The independent-defect hypothesis~\cite{delCampo2018,
GomezRuiz2020} provides a clear organizing principle for defect statistics, but our results identify the dynamical assumption on which it relies. In the integrable models~\cite{delCampo2018, Bando2020, Kiss2025} underlying previous full-counting-statistics predictions, the evolution factorizes into
independent quasiparticle sectors: their occupations do not scatter, and the counting-statistics generating function retains its Poisson-binomial form. The non-integrable Rydberg Hamiltonian lacks this protection. Defects generated during the critical passage can subsequently rearrange and develop correlations through interacting post-critical dynamics. Our results therefore separate two notions that need not coincide: Kibble-Zurek scaling of the mean defect density and statistical independence of the defects. The former can remain approximately valid while the latter fails, as revealed by the super-Poissonian variance and by the defect-correlation pattern that is not inherited from density correlations.

The hold protocol further separates the correlations established during the ramp from subsequent evolution under the final, time-independent Hamiltonian.  By halting energy injection while maintaining a coherent drive, we separate correlations established during the ramp from their subsequent evolution and observe no secular growth over the experimentally accessible hold times, in sharp contrast to diffusive growth reported in two dimensions on similar platforms~\cite{Manovitz2024}.  One possible origin of this arrested dynamics is an emergent prethermal constraint deep in the ordered phase~\cite{Berges:2004,Moeckel:2008,Mori:2018, Darbha2025}. Processes that substantially rearrange or annihilate domain walls may require off-resonant virtual excursions outside the low-energy ordered manifold, suppressing their effective matrix elements and generating a long relaxation time. In two dimensions, the larger coordination number and the many local paths available for deforming an extended domain wall may provide additional resonant channels and weaken this kinetic bottleneck, consistent with the coarsening observed in Ref.~\cite{Manovitz2024}. Because the arrest also occurs in disorder-free simulations, positional disorder cannot be its primary origin, although it may further slow transport and dephase coherent oscillations in the experiment~\cite{Dag2025}. Varying the final hold point, geometry, disorder strength, and ramp and hold durations, while measuring the signed defect correlator and higher cumulants, can distinguish these mechanisms. More generally, our results distinguish defect production during critical freeze-out from subsequent interacting defect dynamics: late-time counting statistics can retain a substantial post-critical imprint and therefore need not provide a direct measure of the Kibble-Zurek freeze-out process alone.

\begin{acknowledgments}

We acknowledge helpful discussions with Michael Kolodrubetz and Fernando G{\'o}mez-Ruiz. This research was supported by the U.S. Department of Energy (DOE) under Contract No. DE-AC02-05CH11231, through the National Energy Research Scientific Computing Center (NERSC), an Office of Science User Facility located at Lawrence Berkeley National Laboratory.

\end{acknowledgments}

\noindent\textbf{Author Contributions.—}
J.B. performed the experiments, data analysis, and error analysis.
A.K. performed the main numerical simulations, numerical analysis, and computational implementation, with preliminary numerical work performed by O.A.A.
M.K. and F.L. conceived the project.
M.K., P.L.S.L., A.K., and S.D. performed the conceptual analysis.
M.K., K.K., and D.C. coordinated the collaboration.
E.R., S.-T.W., K.K., and D.C. supervised the project.
M.K., P.L.S.L., A.K, and S.D. co-wrote the manuscript with revisions from all authors.



\bibliography{references.bib}

@article{Bando2020,
  author  = {Bando, Y. and Nishimori, H. and Albash, T. and Lidar, D. A. and Suzuki, S. and del Campo, A.},
  title   = {Probing the universality of topological defect formation in a quantum annealer: {K}ibble--{Z}urek mechanism and beyond},
  journal = {Physical Review Research},
  volume  = {2},
  number  = {3},
  pages   = {033369},
  year    = {2020},
  doi     = {10.1103/PhysRevResearch.2.033369}
}

@article{Bernien2017,
  author  = {Bernien, H. and Schwartz, S. and Keesling, A. and Levine, H. and Omran, A. and Pichler, H. and Choi, S. and Zibrov, A. S. and Endres, M. and Greiner, M. and Vuleti{\'c}, V. and Lukin, M. D.},
  title   = {Probing many-body dynamics on a 51-atom quantum simulator},
  journal = {Nature},
  volume  = {551},
  number  = {7682},
  pages   = {579--584},
  year    = {2017},
  doi     = {10.1038/nature24622}
}

@misc{bloqadepython:2024,
  author       = {Weinberg, Phillip and Wu, Kai-Hsin and John Long and Luo, Xiu-zhe (Roger)},
  title        = {QuEraComputing/bloqade-python: v0.15.11},
  month        = may,
  year         = 2024,
  publisher    = {Zenodo},
  version      = {v0.15.11},
  doi          = {10.5281/zenodo.11114110},
  url          = {https://doi.org/10.5281/zenodo.11114110}
}

@misc{BloqadeJulia:2023,
  url = {https://github.com/QuEraComputing/Bloqade.jl/},
  title = {Bloqade.jl: {P}ackage for the quantum computation and quantum simulation based on the neutral-atom architecture.},
  year = {2023}
}

@misc{SM,
  url = {},
  title = {See {S}upplemental {M}aterial [url] for details of readout error mitigation and error estimation procedures and the additional numerical results, which includes {R}efs.~\cite{Majumdar:2023, Bravyi2021, perlmutter:2023}},
  year={}
  }

@article{Bray1994,
  author  = {Bray, A. J.},
  title   = {Theory of phase-ordering kinetics},
  journal = {Advances in Physics},
  volume  = {43},
  number  = {3},
  pages   = {357--459},
  year    = {1994},
  doi     = {10.1080/00018739400101505}
}

@article{Bray2003,
  author  = {Bray, A. J.},
  title   = {Coarsening dynamics of phase-separating systems},
  journal = {Philosophical Transactions of the Royal Society A},
  volume  = {361},
  number  = {1805},
  pages   = {781--791},
  year    = {2003},
  doi     = {10.1098/rsta.2002.1164}
}

@article{Chuang1991,
  author  = {Chuang, I. and Durrer, R. and Turok, N. and Yurke, B.},
  title   = {Cosmology in the laboratory: defect dynamics in liquid crystals},
  journal = {Science},
  volume  = {251},
  number  = {4999},
  pages   = {1336--1342},
  year    = {1991},
  doi     = {10.1126/science.251.4999.1336}
}

@article{Cui2020,
  author  = {Cui, Jin-Ming and Gómez-Ruiz, Fernando J. and Huang, Yun-Feng and Li, Chuan-Feng and Guo, Guang-Can and del Campo, Adolfo},
  title   = {Experimentally testing quantum critical dynamics beyond the {K}ibble–{Z}urek mechanism},
  journal = {Communications Physics},
  volume  = {3},
  pages   = {44},
  year    = {2020},
  doi     = {10.1038/s42005-020-0306-6}
}

@misc{Dag2025,
      title={Emergent disorder and sub-ballistic dynamics in quantum simulations of the {I}sing model using {R}ydberg atom arrays}, 
      author={Ceren B. Dag and Hanzhen Ma and P. Myles Eugenio and Fang Fang and Susanne F. Yelin},
      year={2025},
      eprint={2411.13643},
      archivePrefix={arXiv},
      primaryClass={quant-ph},
      url={https://arxiv.org/abs/2411.13643}, 
}

@misc{Darbha2025,
      title={Probing emergent prethermal dynamics and resonant melting on a programmable quantum simulator}, 
      author={Siva Darbha and Alexey Khudorozhkov and Pedro L. S. Lopes and Fangli Liu and Ermal Rrapaj and Jan Balewski and Majd Hamdan and Pavel E. Dolgirev and Alexander Schuckert and Katherine Klymko and Sheng-Tao Wang and Mikhail D. Lukin and Daan Camps and Milan Kornjača},
      year={2025},
      eprint={2510.11706},
      archivePrefix={arXiv},
      primaryClass={quant-ph},
      url={https://arxiv.org/abs/2510.11706}, 
}

@article{delCampo2018,
  author  = {del Campo, Adolfo},
  title   = {Universal Statistics of Topological Defects Formed in a Quantum Phase Transition},
  journal = {Physical Review Letters},
  volume  = {121},
  number  = {20},
  pages   = {200601},
  year    = {2018},
  doi     = {10.1103/PhysRevLett.121.200601}
}

@article{Rysti2021,
   title={Suppressing the Kibble-Zurek Mechanism by a Symmetry-Violating Bias},
   volume={127},
   ISSN={1079-7114},
   url={http://dx.doi.org/10.1103/PhysRevLett.127.115702},
   DOI={10.1103/physrevlett.127.115702},
   number={11},
   journal={Physical Review Letters},
   publisher={American Physical Society (APS)},
   author={Rysti, J. and Mäkinen, J. T. and Autti, S. and Kamppinen, T. and Volovik, G. E. and Eltsov, V. B.},
   year={2021},
   month=sep }

@article{Zhang2025b,
   title={Observation of Near-Critical Kibble-Zurek Scaling in Rydberg Atom Arrays},
   volume={135},
   ISSN={1079-7114},
   url={http://dx.doi.org/10.1103/2gwz-65w1},
   DOI={10.1103/2gwz-65w1},
   number={9},
   journal={Physical Review Letters},
   publisher={American Physical Society (APS)},
   author={Zhang, Tao and Wang, Hanteng and Zhang, Wenjun and Wang, Yuqing and Du, Angrui and Li, Ziqi and Wu, Yujia and Li, Chengshu and Hu, Jiazhong and Zhai, Hui and Chen, Wenlan},
   year={2025},
   month=aug }

@article{Mori:2018,
doi = {10.1088/1361-6455/aabcdf},
url = {https://dx.doi.org/10.1088/1361-6455/aabcdf},
year = {2018},
month = {may},
publisher = {IOP Publishing},
volume = {51},
number = {11},
pages = {112001},
author = {Mori, Takashi and Ikeda, Tatsuhiko N and Kaminishi, Eriko and Ueda, Masahito},
title = {Thermalization and prethermalization in isolated quantum systems: a theoretical overview},
journal = {Journal of Physics B: Atomic, Molecular and Optical Physics}
}

@article{Moeckel:2008,
  title = {Interaction Quench in the Hubbard Model},
  author = {Moeckel, Michael and Kehrein, Stefan},
  journal = {Phys. Rev. Lett.},
  volume = {100},
  issue = {17},
  pages = {175702},
  numpages = {4},
  year = {2008},
  month = {May},
  publisher = {American Physical Society},
  doi = {10.1103/PhysRevLett.100.175702},
  url = {https://link.aps.org/doi/10.1103/PhysRevLett.100.175702}
}

@article{Berges:2004,
  title = {Prethermalization},
  author = {Berges, J. and Bors\'anyi, Sz. and Wetterich, C.},
  journal = {Phys. Rev. Lett.},
  volume = {93},
  issue = {14},
  pages = {142002},
  numpages = {4},
  year = {2004},
  month = {Sep},
  publisher = {American Physical Society},
  doi = {10.1103/PhysRevLett.93.142002},
  url = {https://link.aps.org/doi/10.1103/PhysRevLett.93.142002}
}

@article{Bravyi2021,
  title = {Mitigating measurement errors in multiqubit experiments},
  author = {Bravyi, Sergey and Sheldon, Sarah and Kandala, Abhinav and Mckay, David C. and Gambetta, Jay M.},
  journal = {Phys. Rev. A},
  volume = {103},
  issue = {4},
  pages = {042605},
  numpages = {12},
  year = {2021},
  month = {Apr},
  publisher = {American Physical Society},
  doi = {10.1103/PhysRevA.103.042605},
  url = {https://link.aps.org/doi/10.1103/PhysRevA.103.042605}
}

@article{delCampo2021,
  author  = {del Campo, Adolfo and Gómez‐Ruiz, Fernando J. and Li, Zhi-Hong and Xia, Chuan-Yin and Zeng, Hua-Bi and Zhang, Hai-Qing},
  title   = {Universal statistics of vortices in a newborn holographic superconductor: beyond the {K}ibble–{Z}urek mechanism},
  journal = {Journal of High Energy Physics},
  volume  = {2021},
  number  = {6},
  pages   = {61},
  year    = {2021},
  doi     = {10.1007/JHEP06(2021)061}
}

@article{delCampo2022,
  title = {Locality of spontaneous symmetry breaking and universal spacing distribution of topological defects formed across a phase transition},
  author = {del Campo, Adolfo and G\'omez-Ruiz, Fernando Javier and Zhang, Hai-Qing},
  journal = {Phys. Rev. B},
  volume = {106},
  issue = {14},
  pages = {L140101},
  numpages = {5},
  year = {2022},
  month = {Oct},
  publisher = {American Physical Society},
  doi = {10.1103/PhysRevB.106.L140101},
  url = {https://link.aps.org/doi/10.1103/PhysRevB.106.L140101}
}

@article{delCampo2023,
  author  = {del Campo, Adolfo and Zeng, Hui-Bo and Xia, Cheng-Yun and Li, Zu-Huan},
  title   = {Universal Breakdown of {K}ibble--{Z}urek Scaling in Fast Quenches across a Phase Transition},
  journal = {Physical Review Letters},
  volume  = {130},
  number  = {6},
  pages   = {060402},
  year    = {2023},
  doi     = {10.1103/PhysRevLett.130.060402}
}

@article{Du2023,
  author  = {Du, Kai and Fang, Xiaochen and Won, Choongjae and De, Chandan and Huang, Fei-Ting and Xu, Wenqian and You, Hoydoo and Gómez-Ruiz, Fernando J. and del Campo, Adolfo and Cheong, Sang-Wook},
  title   = {{K}ibble–{Z}urek mechanism of {I}sing domains},
  journal = {Nature Physics},
  volume  = {19},
  pages   = {1495--1501},
  year    = {2023},
  doi     = {10.1038/s41567-023-02112-5}
}

@article{Dziarmaga2010,
   title={Dynamics of a quantum phase transition and relaxation to a steady state},
   volume={59},
   ISSN={1460-6976},
   url={http://dx.doi.org/10.1080/00018732.2010.514702},
   DOI={10.1080/00018732.2010.514702},
   number={6},
   journal={Advances in Physics},
   publisher={Informa UK Limited},
   author={Dziarmaga, Jacek},
   year={2010},
   month=sep, pages={1063–1189}
}

@article{Ebadi2021,
  author  = {Ebadi, S. and Wang, T. T. and Levine, H. and Keesling, A. and Semeghini, G. and Omran, A. and Bluvstein, D. and Samajdar, R. and Pichler, H. and Ho, W. W. and Choi, S. and Sachdev, S. and Greiner, M. and Vuleti{\'c}, V. and Lukin, M. D.},
  title   = {Quantum phases of matter on a 256-atom programmable quantum simulator},
  journal = {Nature},
  volume  = {595},
  pages   = {227--232},
  year    = {2021},
  doi     = {10.1038/s41586-021-03582-4}
}

@article{Gherardini2024,
  title = {Universal Defects Statistics with Strong Long-Range Interactions},
  author = {Gherardini, Stefano and Buffoni, Lorenzo and Defenu, Nicol\`o},
  journal = {Phys. Rev. Lett.},
  volume = {133},
  issue = {11},
  pages = {113401},
  numpages = {7},
  year = {2024},
  month = {Sep},
  publisher = {American Physical Society},
  doi = {10.1103/PhysRevLett.133.113401},
  url = {https://link.aps.org/doi/10.1103/PhysRevLett.133.113401}
}

@article{GomezRuiz2020,
  author  = {G{\'o}mez-Ruiz, Fernando J. and Mayo, Jack J. and del Campo, Adolfo},
  title   = {Full Counting Statistics of Topological Defects after Crossing a Phase Transition},
  journal = {Physical Review Letters},
  volume  = {124},
  number  = {24},
  pages   = {240602},
  year    = {2020},
  doi     = {10.1103/PhysRevLett.124.240602}
}

@article{GomezRuiz2022,
  author  = {G{\'o}mez-Ruiz, Fernando J. and Subires, David and del Campo, Adolfo},
  title   = {Role of boundary conditions in the full counting statistics of topological defects after crossing a continuous phase transition},
  journal = {Physical Review B},
  volume  = {106},
  number  = {13},
  pages   = {134302},
  year    = {2022},
  doi     = {10.1103/PhysRevB.106.134302}
}

@misc{Hirsbrunner2025,
      title={Quantum criticality and nonequilibrium dynamics on a {L}ieb lattice of {R}ydberg atoms}, 
      author={Mark R. Hirsbrunner and Milan Kornjača and Rhine Samajdar and Siva Darbha and Majd Hamdan and Jan Balewski and Ermal Rrapaj and Sheng-Tao Wang and Daan Camps and Fangli Liu and Pedro L. S. Lopes and Katherine Klymko},
      year={2025},
      eprint={2508.05737},
      archivePrefix={arXiv},
      primaryClass={cond-mat.quant-gas},
      url={https://arxiv.org/abs/2508.05737}, 
}

@article{Keesling2019,
  author  = {Keesling, A. and Omran, A. and Levine, H. and Bernien, H. and Pichler, H. and Choi, S. and Samajdar, R. and Schwartz, S. and Silvi, P. and Sachdev, S. and Zoller, P. and Endres, M. and Greiner, M. and Vuleti{\'c}, V. and Lukin, M. D.},
  title   = {Quantum {K}ibble--{Z}urek mechanism and critical dynamics on a programmable {R}ydberg simulator},
  journal = {Nature},
  volume  = {568},
  pages   = {207--211},
  year    = {2019},
  doi     = {10.1038/s41586-019-1070-1}
}

@article{Kibble1976,
  author  = {Kibble, T. W. B.},
  title   = {Topology of cosmic domains and strings},
  journal = {Journal of Physics A: Mathematical and General},
  volume  = {9},
  number  = {8},
  pages   = {1387--1398},
  year    = {1976},
  doi     = {10.1088/0305-4470/9/8/029}
}

@article{Kiss2025,
  author  = {Kiss, O. and Teplitskiy, D. and Grossi, M. and Mandarino, A.},
  title   = {Statistics of topological defects across a phase transition in a superconducting quantum processor},
  journal = {Quantum Science and Technology},
  volume  = {10},
  number  = {3},
  pages   = {035037},
  year    = {2025},
  doi     = {10.1088/2058-9565/addf75}
}

@article{Kolodrubetz2012a,
  title = {Nonequilibrium dynamics of bosonic {M}ott insulators in an electric field},
  author = {Kolodrubetz, M. and Pekker, D. and Clark, B. K. and Sengupta, K.},
  journal = {Phys. Rev. B},
  volume = {85},
  issue = {10},
  pages = {100505},
  numpages = {5},
  year = {2012},
  month = {Mar},
  publisher = {American Physical Society},
  doi = {10.1103/PhysRevB.85.100505},
  url = {https://link.aps.org/doi/10.1103/PhysRevB.85.100505}
}

@article{Kolodrubetz2012b,
  title = {Nonequilibrium Dynamic Critical Scaling of the Quantum {I}sing Chain},
  author = {Kolodrubetz, Michael and Clark, Bryan K. and Huse, David A.},
  journal = {Phys. Rev. Lett.},
  volume = {109},
  issue = {1},
  pages = {015701},
  numpages = {5},
  year = {2012},
  month = {Jul},
  publisher = {American Physical Society},
  doi = {10.1103/PhysRevLett.109.015701},
  url = {https://link.aps.org/doi/10.1103/PhysRevLett.109.015701}
}

@misc{Majumdar:2023,
      title={Best practices for quantum error mitigation with digital zero-noise extrapolation}, 
      author={Ritajit Majumdar and Pedro Rivero and Friederike Metz and Areeq Hasan and Derek S Wang},
      year={2023},
      eprint={2307.05203},
      archivePrefix={arXiv},
      primaryClass={quant-ph},
      url={https://arxiv.org/abs/2307.05203}, 
}

@article{Manovitz2024,
  author  = {Manovitz, T. and Li, S. H. and Ebadi, S. and Samajdar, R. and Geim, A. A. and Evered, S. J. and Bluvstein, D. and Zhou, H. and Koyluoglu, N. U. and Feldmeier, J. and Dolgirev, P. E. and Maskara, N. and Kalinowski, M. and Sachdev, S. and Huse, D. A. and Greiner, M. and Vuleti{\'c}, V. and Lukin, M. D.},
  title   = {Quantum coarsening and collective dynamics on a programmable quantum simulator},
  journal = {Nature},
  volume  = {635},
  pages   = {278--283},
  year    = {2024},
  doi     = {10.1038/s41586-024-08353-5}
}

@article{Mattioli2025,
  author  = {Mattioli, M. and Defenu, N. and Ricci, C. and Ruffo, S.},
  title   = {Defect Production across Higher-Order Phase Transitions beyond {K}ibble--{Z}urek},
  journal = {Physical Review Letters},
  volume  = {134},
  number  = {1},
  pages   = {010409},
  year    = {2025},
  doi     = {10.1103/PhysRevLett.134.010409}
}

@article{Monaco2009,
  author  = {Monaco, R. and Mygind, J. and Rivers, R. J. and Koshelets, V. P.},
  title   = {Spontaneous fluxoid formation in superconducting loops},
  journal = {Physical Review B},
  volume  = {80},
  number  = {18},
  pages   = {180501},
  year    = {2009},
  doi     = {10.1103/PhysRevB.80.180501}
}

@misc{perlmutter:2023,
  url = {https://www.nersc.gov/systems/perlmutter},
  title = {Perlmutter.},
  year = {2023}
}

@article{Pyka2013,
  author  = {Pyka, K. and Keller, J. and Partner, H. L. and Nigmatullin, R. and Burgermeister, T. and Meier, D. M. and Kuhlmann, K. and Retzker, A. and Plenio, M. B. and Zoller, P. and Schmidt-Kaler, F. and Poschinger, U. G.},
  title   = {Topological defect formation and spontaneous symmetry breaking in ion {C}oulomb crystals},
  journal = {Nature Communications},
  volume  = {4},
  pages   = {2291},
  year    = {2013},
  doi     = {10.1038/ncomms3291}
}

@article{Roychowdhury2021,
  author  = {Roychowdhury, K. and Moessner, R. and Das, A.},
  title   = {Dynamics and correlations at a quantum phase transition beyond {K}ibble--{Z}urek},
  journal = {Physical Review B},
  volume  = {104},
  number  = {1},
  pages   = {014406},
  year    = {2021},
  doi     = {10.1103/PhysRevB.104.014406}
}

@misc{Samajdar2024,
      title={Quantum and classical coarsening and their interplay with the {K}ibble-{Z}urek mechanism}, 
      author={Rhine Samajdar and David A. Huse},
      year={2024},
      eprint={2401.15144},
      archivePrefix={arXiv},
      primaryClass={quant-ph},
      url={https://arxiv.org/abs/2401.15144}, 
}

@article{Ulm2013,
  author  = {Ulm, S. and Roßnagel, J. and Jacob, G. and Degünther, C. and Dawkins, S. T. and Poschinger, U. G. and Nigmatullin, R. and Retzker, A. and Plenio, M. B. and Schmidt-Kaler, F. and Singer, K.},
  title   = {Observation of the {K}ibble--{Z}urek scaling law for defect formation in ion crystals},
  journal = {Nature Communications},
  volume  = {4},
  pages   = {2290},
  year    = {2013},
  doi     = {10.1038/ncomms2290}
}

@article{Weiler2008,
  author  = {Weiler, C. N. and Neely, T. W. and Scherer, D. R. and Bradley, A. S. and Davis, M. J. and Anderson, B. P.},
  title   = {Spontaneous vortices in the formation of {Bose--Einstein} condensates},
  journal = {Nature},
  volume  = {455},
  number  = {7215},
  pages   = {948--951},
  year    = {2008},
  doi     = {10.1038/nature07334}
}

@misc{Wurtz2023,
      title={Aquila: QuEra's 256-qubit neutral-atom quantum computer}, 
      author={Jonathan Wurtz and Alexei Bylinskii and Boris Braverman and Jesse Amato-Grill and Sergio H. Cantu and Florian Huber and Alexander Lukin and Fangli Liu and Phillip Weinberg and John Long and Sheng-Tao Wang and Nathan Gemelke and Alexander Keesling},
      year={2023},
      eprint={2306.11727},
      archivePrefix={arXiv},
      primaryClass={quant-ph}
}

@article{Xu2016,
  author  = {Xu, K. and Zhang, J. and Peng, X. and Rajendran, N. and Chu, J. and Du, J.},
  title   = {Simulating the {K}ibble--{Z}urek mechanism of the {I}sing model with a superconducting qubit system},
  journal = {Scientific Reports},
  volume  = {6},
  pages   = {22667},
  year    = {2016},
  doi     = {10.1038/srep22667}
}

@article{Zurek1985,
  author  = {Zurek, W. H.},
  title   = {Cosmological experiments in superfluid helium?},
  journal = {Nature},
  volume  = {317},
  number  = {6037},
  pages   = {505--508},
  year    = {1985},
  doi     = {10.1038/317505a0}
}

@book{Sachdev2011,
  title={Quantum Phase Transitions},
  author={Subir Sachdev},
  publisher={Cambridge University Press},
  year={2011},
  month={4},
  edition={2nd},
  isbn={978-0-511-97376-5},
  doi={10.1017/CBO9780511973765}
}

@article{Sengupta2004,
  title = {Quench dynamics across quantum critical points},
  author = {Sengupta, K. and Powell, Stephen and Sachdev, Subir},
  journal = {Phys. Rev. A},
  volume = {69},
  issue = {5},
  pages = {053616},
  numpages = {10},
  year = {2004},
  month = {May},
  publisher = {American Physical Society},
  doi = {10.1103/PhysRevA.69.053616},
  url = {https://link.aps.org/doi/10.1103/PhysRevA.69.053616}
}

@Article{Zhang2025,
author={Zhang, Jin and Cant{\'u}, Sergio H. and Liu, Fangli and Bylinskii, Alexei and Braverman, Boris and Huber, Florian and Amato-Grill, Jesse and Lukin, Alexander and Gemelke, Nathan and Keesling, Alexander and Wang, Sheng-Tao and Meurice, Yannick and Tsai, Shan-Wen},
title={Probing quantum floating phases in {R}ydberg atom arrays},
journal={Nature Communications},
year={2025},
month={Jan},
day={16},
volume={16},
number={1},
pages={712},
issn={2041-1723},
doi={10.1038/s41467-025-55947-2},
url={https://doi.org/10.1038/s41467-025-55947-2}
}

\clearpage
\onecolumngrid
\begin{center}
{\bf Supplemental Material: ``Observation of an anomaly in the statistics of Kibble-Zurek defects''}
\end{center}
\setcounter{page}{1}
\thispagestyle{empty}
\setcounter{secnumdepth}{3}
\renewcommand\thesection{S\arabic{section}}

In this Supplemental Material, we present the details of readout error mitigation and error estimation procedures in Sec.~\ref{sec:mitigation}. The additional numerical results follow, including the calculation details (Sec.~\ref{sec:numerics}), the effect of the perfect blockade constraint (Sec.~\ref{sec:subspace}), details of defect correlations and statistics (Sec.~\ref{sec:corrs}), and supplementary hold dynamics results presenting the spectrum of defect number oscillations (Sec.~\ref{sec:hold}).

\section{Readout Error Mitigation}
\label{sec:mitigation}

The measurement of defect counting statistics in one-dimensional Rydberg atom chains relies on accurately determining the state of each atom at the end of the evolution, as the defect number is sensitive to readout errors. In our experiments, the dominant sources of readout errors are finite detection fidelity during fluorescence imaging and Rydberg state decay during the comparatively long imaging time~\cite{Wurtz2023}, which manifest as state-dependent bit-flip errors. Each experimental shot yields a fluorescence image of the atom chain, which is thresholded to produce a binary string identifying ground ($\ket{0}$) and excited ($\ket{1}$) states~\cite{Wurtz2023}. From these measured bitstrings, we extract defect-relevant observables by counting domain wall defects, which correspond to ``00'' and ``11'' patterns. Any readout error can introduce these patterns, leading to a systematically increased domain wall count and a biased variance.

\begin{figure}[htb]
\centering
\includegraphics[width=0.9\columnwidth]{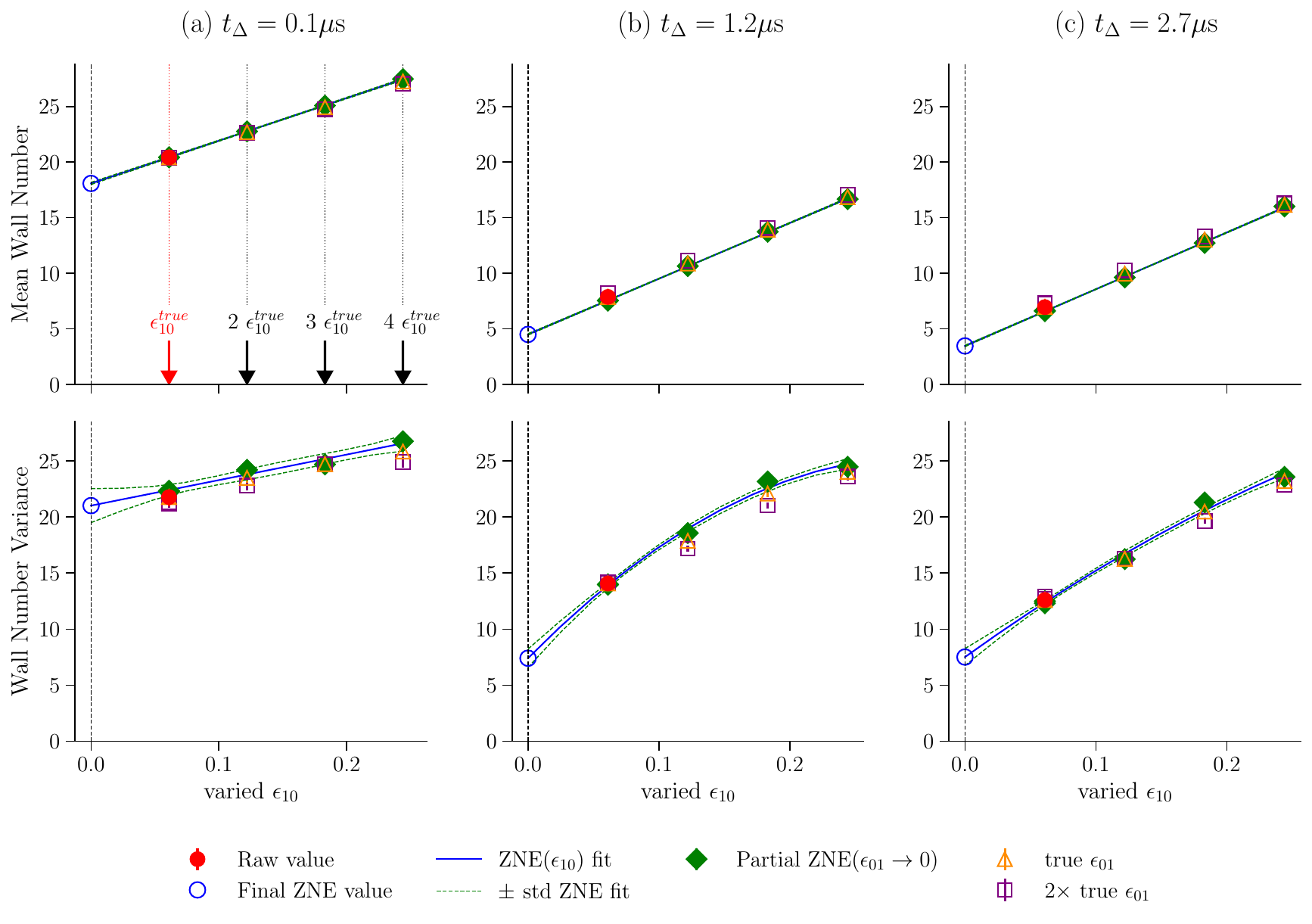}
\caption{
\textbf{Zero-noise extrapolation (ZNE) for readout error mitigation.}
The two-step ZNE procedure is shown for the mean wall number (top row) and wall number variance (bottom row) at three different ramp times $t_\Delta$ (columns a-c).
The process starts with raw experimental data (red circle at $\epsilon_{10}^{\text{true}}$) and data points where noise has been artificially amplified (orange triangles and purple squares).
First, a linear extrapolation in the $\epsilon_{01}$ error is performed for fixed values of $\epsilon_{10}$ to obtain the partially-mitigated intermediate data (green diamonds).
Second, these intermediate points are extrapolated to the zero-noise limit ($\epsilon_{10} \to 0$).
A linear model (blue line) is sufficient for the mean wall number, while a quadratic model is required for the variance.
The final, fully error-mitigated value is the extrapolated point at $\epsilon_{10}=0$ (blue open circle).
The dashed green lines represent the one-standard-deviation confidence interval of the fit.
}
\label{fig:ZNE}
\end{figure}

To mitigate the impact of readout errors, we employ zero-noise extrapolation (ZNE). While ZNE is commonly used in quantum computing to suppress gate errors, a process that typically requires deliberately adding noise and performing additional experiments~\cite{Majumdar:2023}, our protocol is tailored to exclusively correct readout errors. This allows us to perform ZNE entirely as a post-processing step on the measured data, without any additional experimental overhead, as long as the readout error probabilities are reliably and independently inferred from dedicated calibration     experiments.

To account for potential hardware drift over data-taking sessions lasting several hours, we calibrate the readout errors both before and after the main experiment. Using the standard method of fitting damped Rabi oscillations~\cite{Wurtz2023}, we measure the probability of misclassifying a true Rydberg state as ground ($\epsilon_{10}$) and a true ground state as Rydberg ($\epsilon_{01}$). The average values over the  calibrations passes were $\epsilon_{10}^{\text{true}} = 0.061 \pm 0.004$ and $\epsilon_{01}^{\text{true}} = 0.009 \pm 0.002$, with the uncertainties reflecting the difference between the passes. Qubits are spatially isolated and identical atoms that are imaged simultaneously; thus, no higher-order correlated readout error channels appear, making our two-parameter readout error model complete.

We implement ZNE in the two-dimensional space spanned by the error parameters $(\epsilon_{01}, \epsilon_{10})$ through noisy resampling. The procedure involves artificially amplifying the noise in the measured bitstrings by randomly flipping bits with probabilities corresponding to multiples of the calibrated error rates, $(\alpha \epsilon_{01}^{\text{true}}, \beta \epsilon_{10}^{\text{true}})$. For each pair of noise amplification factors $(\alpha, \beta)$, we recalculate the average wall number and variance on the resampled bitstring distribution.  For both observables  their statistical errors were computed as well. To reduce the statistical uncertainty, the sampling was performed four times, and the results were averaged.

Since $\epsilon_{01}^{\text{true}}$ is very small, we factorize the two-dimensional extrapolation into two steps, as illustrated in Fig.~\ref{fig:ZNE}. First, for several fixed multiples of $\epsilon_{10}$, we apply a linear extrapolation over  $\epsilon_{01}$ to mitigate the small $\epsilon_{01}^{true}$ error by regressing it to zero. The results of this initial step are shown as green diamonds. Second, using these intermediate points, we assume a model for $\epsilon_{10}$ dependence, fit it, and perform an extrapolation to the $\epsilon_{10} \rightarrow 0$  limit. For the wall number (top row), a linear fit is sufficient across all ramp times $t_\Delta$. For the wall variance (bottom row), the data exhibit clear curvature, calling for a quadratic extrapolation. The final, fully mitigated values are shown as blue circles.

The ZNE models are chosen based on the observed data trends. The linear model for the wall number explains well the $\epsilon_{10}$ dependence for all $t_\Delta$, with the relative ZNE correction increasing with $t_\Delta$ while the extrapolation error remains small. The linear model for wall number is also fully consistent with the commonly employed confusion-matrix-inversion-based readout mitigation~\cite {Bravyi2021}. It is evident that the linear model does not capture the trend observed in the experimentally measured wall variance data shown in the bottom row of Fig.~\ref{fig:ZNE}, particularly at intermediate $t_\Delta$. This observation supports our choice of a quadratic fit model. Furthermore, both a linear ZNE model and confusion matrix inversion for the wall variance fail when tested on synthetic data.

The ZNE fit was performed on observables weighted by their statistical errors; thus, the uncertainty in the extrapolated final values already reflects the statistical uncertainty of the raw bitstring distribution. This uncertainty of the extrapolation is represented by the one-standard-deviation bands (dashed green lines). For the wall variance, the uncertainty grows significantly as $\epsilon_{10} \rightarrow 0$ because the noise model is quadratic in $\epsilon_{10} -\epsilon_{10}^{true} $. This uncertainty is propagated to the final results presented in the main text. 

The ZNE procedure described above yields corrected observable values along with their associated statistical errors. In addition, we estimated the systematic error introduced by the ZNE correction itself. The two input parameters $\epsilon_{10}^{\text{true}}$ and $\epsilon_{01}^{\text{true}}$ determine the output of the ZNE postprocessing, and each is known with an uncertainty $\delta \epsilon_{10}^{\text{true}}$ and $\delta \epsilon_{01}^{\text{true}}$, respectively.
To assess the impact of these uncertainties, we repeated the ZNE estimation process an additional four times, each time assuming $\tilde{\epsilon}_{10}^{\text{true}}$ and $\tilde{\epsilon}_{01}^{\text{true}}$ to be $\epsilon_{10}^{\text{true}} \pm \delta \epsilon_{10}^{\text{true}}$ and $\epsilon_{01}^{\text{true}} \pm \delta \epsilon_{01}^{\text{true}}$, respectively.
For each ramp time $t_\Delta$, we computed the maximum deviation between the baseline ZNE result and the four variant ZNE results obtained with perturbed $\epsilon$ values by  $\pm\delta\epsilon$. These maximum deviations are reported as systematic uncertainties for the final observables and are shown as shaded bands in Fig.~3 of main text.

\section{Supplementary numerical results} 

\subsection{Numerical Methods}
\label{sec:numerics}

The numerical simulations of the Rydberg atom chain we performed using exact diagonalization (ED), implemented in Julia via the open-source package \texttt{Bloqade.jl}~\cite{BloqadeJulia:2023}. All simulations are executed on the Perlmutter supercomputer~\cite{perlmutter:2023} at the National Energy Research Scientific Computing Center (NERSC), hosted at Lawrence Berkeley National Laboratory. Two types of calculations are performed: (i) projected within the Rydberg blockaded Hilbert subspace, allowing access to system sizes up to $L = 24$ atoms, and (ii) in the full Hilbert space, with chains of up to $L = 20$ atoms. This ED-based approach enables us to compute the exact many-body wavefunction and obtain numerically exact time-evolution data for all observables of interest.

The atomic spacing is fixed at $a = 6.2~\mu\mathrm{m}$, corresponding to a dimensionless blockade ratio $R_b / a = 1.35$, where the blockade radius is $R_b = (C_6/\Omega)^{1/6}$, with the van der Waals coefficient $C_6 / 2\pi = 862690 \,\, \mathrm{MHz} \cdot \mu\mathrm{m}^6$ and the maximal Rabi frequency set to $\Omega_\mathrm{max} / 2\pi = 2.5~\mathrm{MHz}$. The Hamiltonian (Eq.~1 of main text) is explicitly time-dependent, with the Rabi frequency $\Omega(t)$ and detuning $\Delta(t)$ temporal profiles the same as in Fig.~2(b) and Fig.~4(a) of main text. In particular, $\Omega(t)$ features a rapid initial ramp-up and final ramp-down of duration $0.5~\mu\mathrm{s}$ each, while $\Delta(t)$ is linearly swept from $\Delta_\mathrm{min} / 2\pi = -2.5~\mathrm{MHz}$ to $\Delta_\mathrm{max} / 2\pi = 4.0~\mathrm{MHz}$ over a total ramp time $t_\Delta \in [0.066, 32.5]~\mu\mathrm{s}$.

All simulations are initialized in the fully unexcited product state $\ket{0}^{\otimes L}$, and only the final state at the end of the evolution is stored for analysis. Observables, including the domain-wall density, its variance, and correlation functions, are computed from this final state. Whenever correlation functions are evaluated, we use the shifted Ising variables, $\tilde{n}_i \equiv 2 \left( n_i - \frac{1}{2} \right), \,
\tilde{D}_i \equiv 2 \left( D_i - \frac{1}{2} \right)$, whose eigenvalues are $\pm 1$. This choice centers the observables around zero and facilitates comparison between different quantities. All reported correlation functions are connected, i.e. computed as $\expval{AB}_\mathrm{c} \equiv \langle AB \rangle - \langle A \rangle \langle B \rangle$, and are shown as absolute values throughout.

\subsection{Effect of the Blockade Subspace}
\label{sec:subspace}

\begin{figure}[htb]
\centering
\includegraphics[scale=0.33]{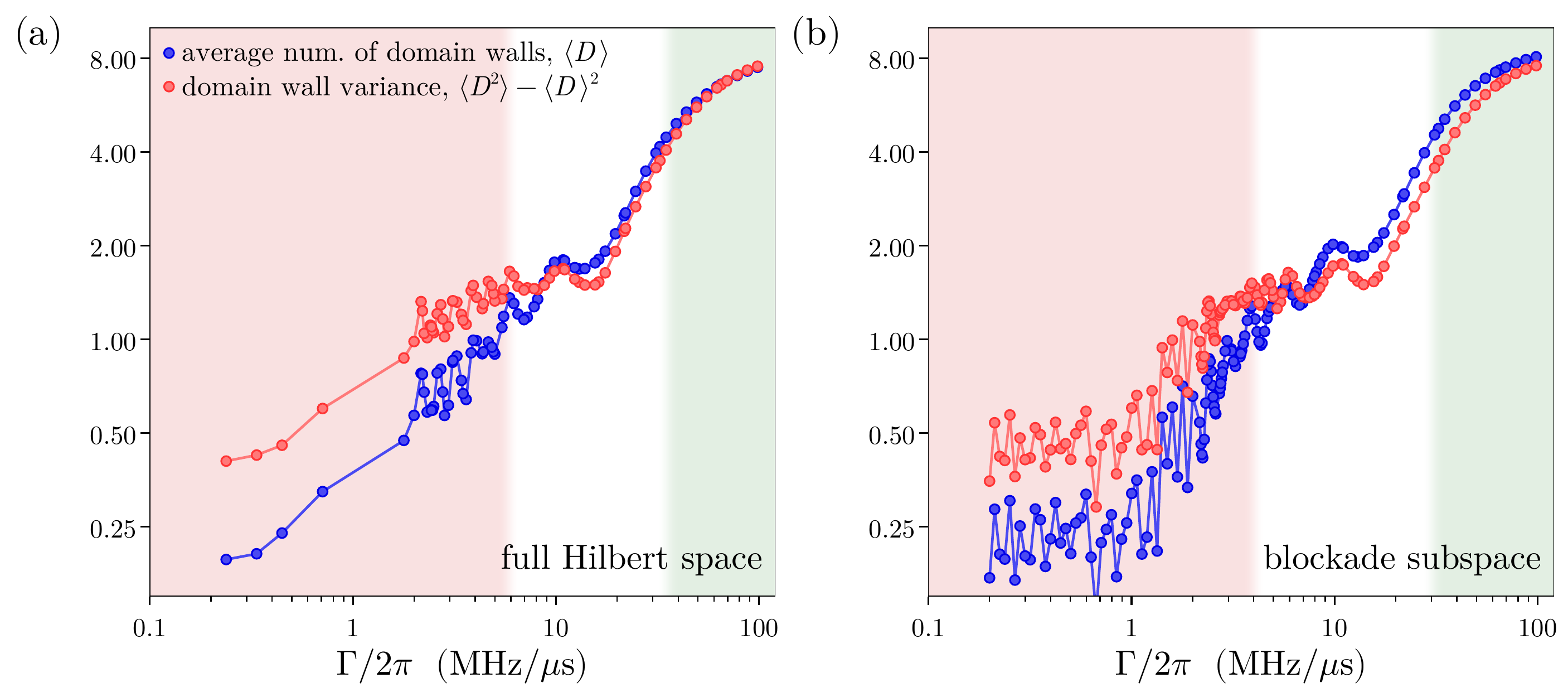}
\caption{\textbf{Effect of using the blockade subspace.} 
Average domain wall number $\langle D \rangle$ (blue) and its variance $\langle D^2 \rangle - \langle D \rangle^2$ (red) for a chain of $L = 20$ atoms with periodic boundary conditions (PBC), obtained via exact diagonalization. The fast ramp regime, defined by $\langle D \rangle \gtrsim L/4$, is shown in green. The slow ramp regime, defined by $\langle D \rangle < 1$, is shown in red.
(a) Results computed in the full $2^L$-dimensional Hilbert space. 
(b) Results restricted to the blockade subspace, where neighboring atoms cannot simultaneously occupy the Rydberg state.
}
\label{fig:fullspace_vs_subspace}
\end{figure}

To assess the role of the Rydberg blockade constraint in the observed dynamics, we compare numerical simulations performed in the full Hilbert space, where blockade violations are allowed, to those restricted to the blockade subspace, where adjacent atoms cannot be simultaneously excited.  Panels~(a) and~(b) in Fig.~\ref{fig:fullspace_vs_subspace} compare these two approaches. We 
show the average number of domain walls $\langle D \rangle$ and its variance $\langle D^2 \rangle - \langle D \rangle^2$ as a function of the effective sweep rate $\Gamma / 2\pi$ for a chain of $L=20$ atoms with periodic boundary conditions (PBC).

The overall behavior of both observables is qualitatively similar in the two cases. For intermediate ramp rates $\Gamma$, the mean number of domain walls and their variance exhibit a sublinear Kibble-Zurek-like scaling, approximately following $\Gamma^{-1/2}$. At low $\Gamma$, both quantities saturate to constant values, with the variance larger than the mean. This consistency across the two Hilbert spaces demonstrates that the Kibble-Zurek statistics anomaly reported in the main text is not an artifact of restricting to the blockade subspace, but a robust feature of the many-body dynamics itself.

Quantitatively, the blockade constraint suppresses large fluctuations by reducing the accessible configuration space, in particular forbidding ``11'' domain wall configurations. Consequently, the variance in the blockade subspace, the numerical result reported in the main text, remains slightly lower than in the full Hilbert-space case, particularly at intermediate and high ramp rates where blockade-violating configurations are energetically suppressed but not entirely forbidden. At least part of the quantitative difference between the experiment and numerics reported in the main text could be attributed to the blockade subspace constraint. These differences underscore the role of kinetic constraints in shaping excitation statistics, while confirming that blockade-violating processes do not qualitatively affect the anomalous scaling behavior.

\subsection{Defect counting statistics and correlations}
\label{sec:corrs}

\begin{figure}[!htb]
\centering
\includegraphics[scale=0.43]{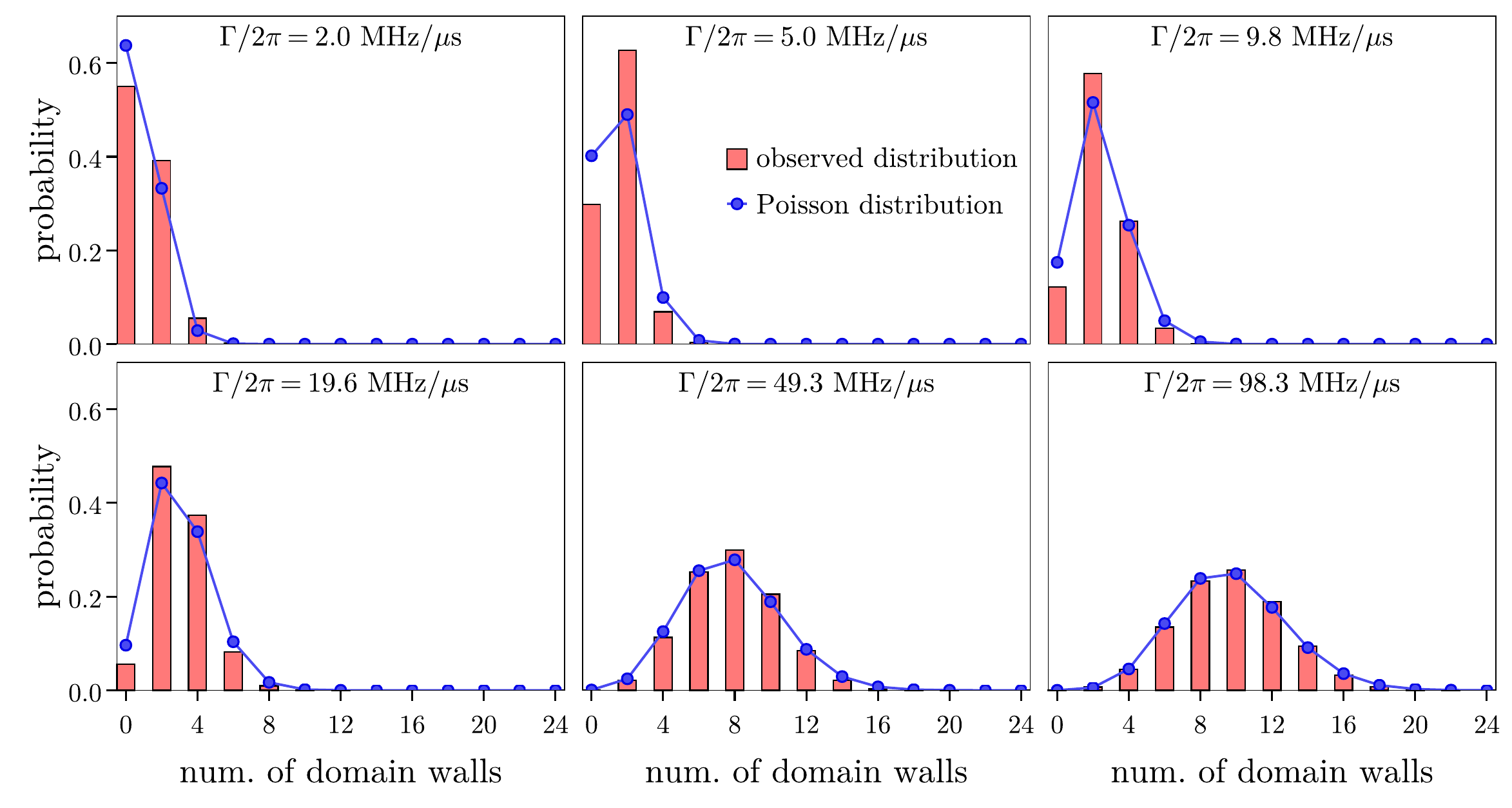}
\caption{\textbf{Defect number distribution.} 
(Red bars) Probability distribution of the number of domain walls for various ramp rates $\Gamma$, obtained numerically via exact diagonalization within the blockade subspace for a chain of $L = 24$ atoms with periodic boundary conditions. (Blue dots and lines) Corresponding normalized (even) Poisson distributions, 
$P(k) = \lambda^k e^{-\lambda} / k!$, 
with $\lambda = \langle D \rangle$ the average number of domain walls. 
Normalization is required since only even values of $k$ are allowed with PBC.
}
\label{fig:ndw_dist}
\end{figure}

\begin{figure}[!htb]
\centering
\includegraphics[scale=0.43]{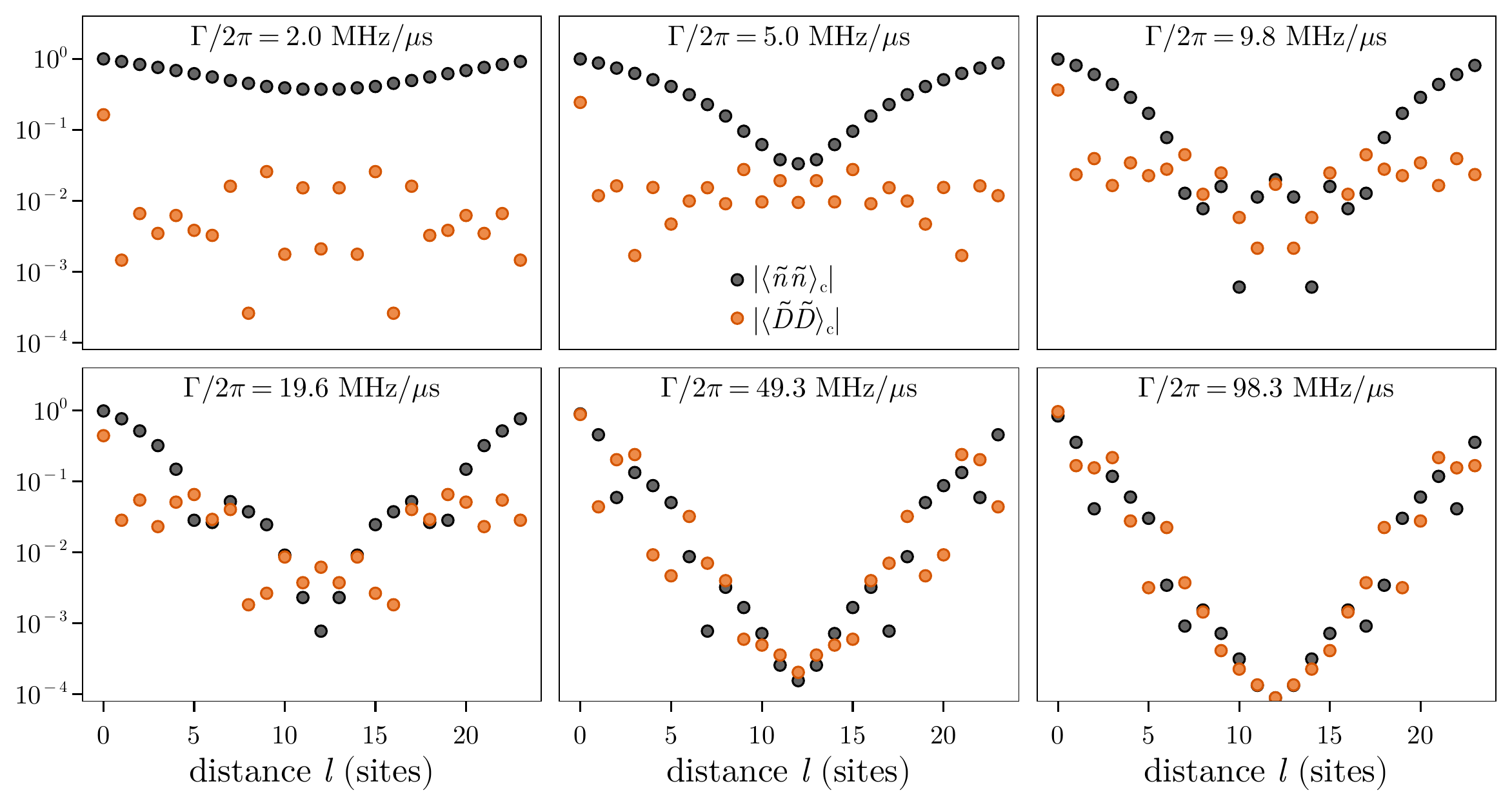}
\caption{\textbf{Correlation functions.} 
Absolute values of the connected correlation functions of the shifted Ising observables 
$\tilde{n}_i = 2(n_i - 1/2)$ (density, black) and 
$\tilde{D}_i = 2(D_i - 1/2)$ (domain wall, orange) 
for various ramp rates $\Gamma$, obtained numerically via exact diagonalization within the blockade subspace for a chain of $L = 24$ atoms with periodic boundary conditions.
}
\label{fig:corr_func}
\end{figure}

To gain further insight into the counting statistics of defects generated during the ramp, we analyze both the distribution of the total number of antiferromagnetic domain walls and their spatial correlations (Figs.~\ref{fig:ndw_dist}--\ref{fig:corr_func}). The probability distributions in Fig.~\ref{fig:ndw_dist} show the likelihood of observing a given number of domain walls for various ramp rates~$\Gamma$, obtained via exact diagonalization within the blockade subspace. For slow ramps, the distributions deviate from the even Poisson form.

As the ramp rate increases, the distributions broaden and gradually approach the Poisson distribution, $P(k \in \mathrm{even}) \propto \lambda^k e^{-\lambda}/k!$, with $\lambda = \langle D \rangle$ the mean number of domain walls. At large $\Gamma$, the observed histograms are approximately described by the even-Poisson reference curves. Note that because periodic boundary conditions enforce that domain walls appear in pairs, only even values of $k$ are allowed, and the corresponding even Poisson distribution is renormalized accordingly.

The change from correlated to uncorrelated defect formation is further supported by the behavior of the connected density–density and domain-wall–domain-wall correlation functions shown in Fig.~\ref{fig:corr_func} (evaluated using the shifted Ising variables defined in the caption). At high~$\Gamma$, the defect correlations decay exponentially with distance, following the correlation length determined by density correlations, reflecting the statistically independent defect formation from correlation-length-sized predomains.  At low~$\Gamma$, defect correlations deviate from density correlations, with the existence of a defect suppressing the formation of nearby defects. The deviations from the Poisson distribution and the correlated nature of the defects for slow ramps are complementary signatures of the defect statistics anomaly we observe. As proposed in the main text, defects created during the critical passage of the fan undergo non-critical dynamics deep in the ordered phase, producing both non-trivial defect correlations and non-Poissonian counting statistics.

\subsection{Hold dynamics}
\label{sec:hold}

\begin{figure}[!htb]
\centering
\includegraphics[scale=0.32]{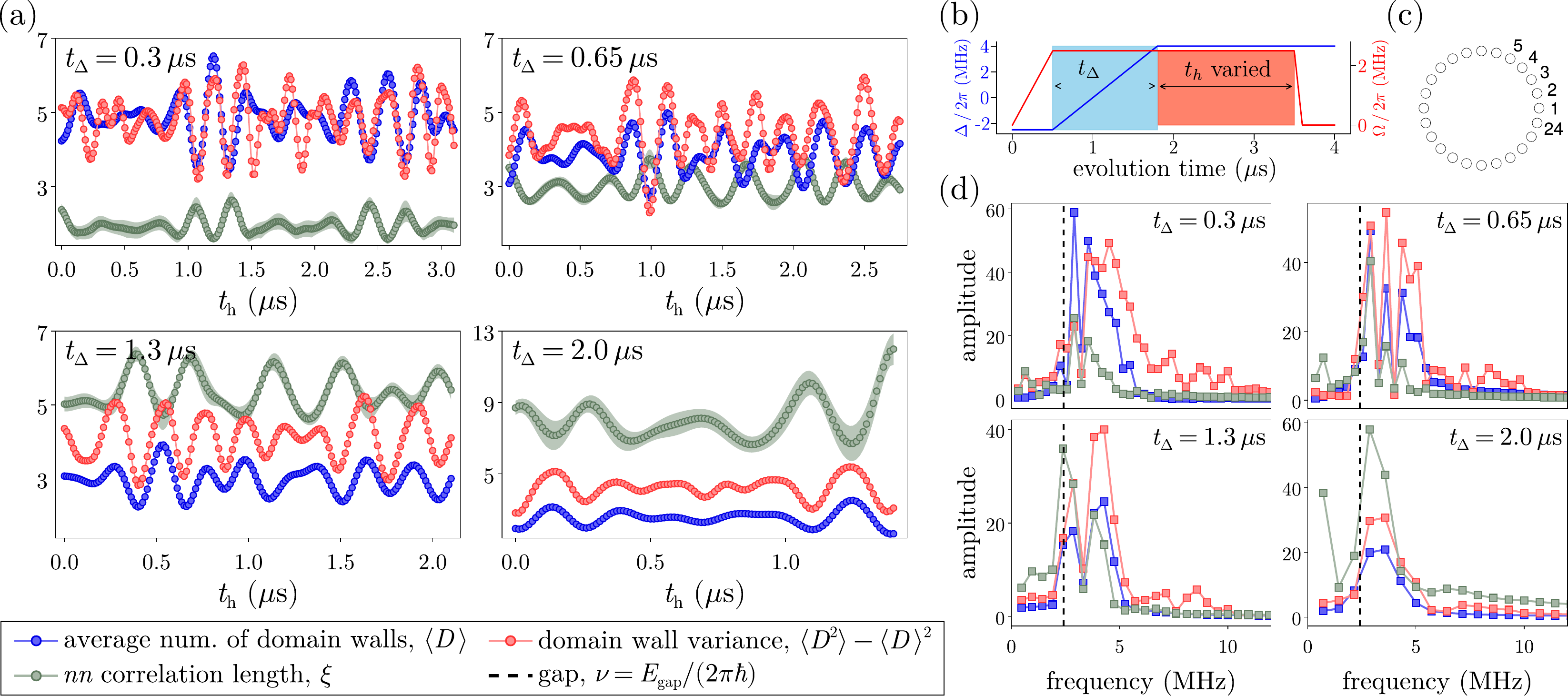}
\caption{ \textbf{Hold dynamics spectrum in simulations.} 
(a) Time evolution of the average number of domain walls $\langle D \rangle$ (blue), its variance $\langle D^2 \rangle - \langle D \rangle^2$ (red), and the density–density correlation length $\xi$ (green) during the hold stage of the experiment, for four different ramp durations $t_{\Delta}$. 
The shaded green band denotes the standard error from fitting the correlation function with an exponential decay. 
(b) Driving fields used in the simulated protocol. 
(c) Geometry of the $L = 24$ atom chain with periodic boundary conditions. 
(d) Fourier transform of the time series in (a); the system’s spectral gap frequency $\nu = E_{\mathrm{gap}} / (2\pi \hbar)$ is indicated by the black dashed line.
}
\label{fig:hold_dw}
\end{figure}

To probe and control the non-critical dynamics in the ordered phase, we perform ``hold'' experiments in which the detuning is kept fixed at its final value after the ramp, and the system evolves under a time-independent Hamiltonian for a variable duration $t_\mathrm{h}$ (Fig.~\ref{fig:hold_dw}). The driving protocol and system geometry used for simulations are shown in Figs.~\ref{fig:hold_dw}(b)–(c). The evolution of the mean domain-wall number $\langle D \rangle$, its variance $\langle D^{2} \rangle - \langle D \rangle^{2}$, and the density–density correlation length~$\xi$ during the hold stage is displayed in Fig.~\ref{fig:hold_dw}(a) for several ramp durations~$t_\Delta$, supplementing the main text results. As reported, all three observables are dominated by their constant running averages, in contrast to the diffusive growth observed in the 2D Rydberg array experiments of Ref.~\cite{Manovitz2024}. Coherent oscillations are superimposed on the main constant signal. The oscillations have well-defined frequencies, suggesting that the system evolves within a subset of low-lying many-body eigenstates.

The origin of the oscillations is manifest in their Fourier spectra, presented in Fig.~\ref{fig:hold_dw}(d). In all cases, the oscillation frequencies below the ground state energy gap (dashed lines) are strongly suppressed, with the dominant frequency coinciding with the first band of excitations starting just after $\nu = E_{\mathrm{gap}}/(2\pi\hbar)$. The gap is calculated from the Hamiltonian at the final detuning (hold Hamiltonian). This demonstrates that the ramp prepares a superposition of the antiferromagnetic ground state and its several low-energy excited states in the ordered phase. The subsequent hold dynamics therefore correspond to coherent amplitude-mode oscillations across the system gap, directly analogous to the Higgs-related oscillations observed in 2D Rydberg arrays \cite{Manovitz2024}. 

\end{document}